\documentclass[12pt]{iopart}

\usepackage{iopams}
\usepackage{graphicx}  

\begin{document}

\title[Motor-motor interaction in models for cargo transport]{Influence of
direct motor-motor interaction in models for cargo transport by a single
team of motors}

\author{Sebasti\'an Bouzat$^{1,2}$ and Fernando Falo$^{1}$}

\address{1 Dpto de F\'{\i}sica de la Materia condensada and BIFI, Universidad de Zaragoza,
50009 Zaragoza, Spain. \\
2 Consejo Nacional de Investigaciones Cient\'{\i}ficas y T\'ecnicas, Centro At\'omico Bariloche (CNEA), (8400) Bariloche, Argentina.}
\ead{bouzat@cab.cnea.gov.ar}

\begin{abstract}
We analyze theoretically the effects of excluded-volume interactions between motors on the 
dynamics of a cargo driven by multiple motors. The model considered 
shares many commons with other recently proposed in the literature, with the addition of 
direct interaction between motors and motor back steps. The cargo is assumed to follow 
a continuum Langevin dynamics, while individual motors evolve following a Monte
Carlo algorithm based on experimentally accessible probabilities for discrete forward and
backward jumps, and attachment and detachment rates. The links between cargo and motors
are considered as non linear springs. By means of numerical simulations we compute the relevant 
quantities characterizing the dynamical properties of the system, and we compare the results to 
those for non interacting motors. We find that interactions
lead to quite relevant changes in the force-velocity relation for cargo, with a
considerable reduction of the stall force, and cause also a notable decrease of the run
length. These effects are mainly due to traffic-like phenomena in the microtubule. The
consideration of several parallel tracks for motors reduces such effects. However, we
find that for realistic values of the number of motors and the number of tracks, the
influence of interactions on the global parameters of transport of cargo are far from being
negligible. Our studies provide also an analysis of the relevance of motor back steps on
the modeling, and of the influence of different assumptions for the detachment rates. In
particular, we discuss these two aspects in connection with the possibility of observing
processive back motion of cargo at large load forces.
\end{abstract}

\vspace{2pc}

\submitto{\PB}
\maketitle


\section{Introduction}

Active transport in cells is mediated by specialized proteins generically called
molecular motors \cite{Howard} which are able to move big cargoes such as vesicles, lipid
droplets or mitochondria \cite{gross2004} at large distances, over long polymer highways
constituted by microtubules or actin bundles. An important behavior of transport on
microtubules is its unidirectional character induced by their anisotropy. One kind of
proteins (kinesins) moves preferently from nucleus to cell periphery whereas other
(dyneins) moves in opposite direction. The precise mechanism by which this translocation
is done has been under active research in biological physics in the last 15 years. Although
some details are still intriguing \cite{BlockBPJ2007}, a reasonable comprehension of how
a single motor works is now available \cite{Howard, Schliwa03,fisher07}.

In last years, new experimental techniques allowed for long time observations with shorter
windows of temporal resolution. As a consequence, it was shown that transport is more
complex than expected from a single motor image: larger processivity lengths than
those expected \cite{grossPNAS2007} and bidirectional motion of some cargoes on ''in vivo'' experiments were
observed \cite{gross2004,welte04}. Both facts are due to the collaboration of more than one motor in
the transport of single cargoes. In particular, bidirectionality is possible due to
the joint action of two different classes of motors. In view of these results, new
experiments and models have flourished \cite{welte04,lipowskyPhysE} to elucidate the details of
transport driven by one \cite{klumpp2005,grossCB2008,mogilner,lipowsky09} and two \cite{tugofwar,lipowskiBioPJ}
teams of motors. However, the understanding of transport by multiple motors is still not as advanced
as that of single motors.


Several realistic models for cargo transport driven by multiple motors specially suitable for kinesin systems in microtubules have been recently developed \cite{lipowskyPhysE,klumpp2005,grossCB2008,mogilner}. Within such context, the effects of direct interactions between motors (i.e. not cargo mediated) have only been analyzed in models that consider equal load sharing \cite{lipowskyPhysE,klumpp2005}. In contrast, studies providing more detailed descriptions of cargo-motor linking and force sharing \cite{grossCB2008,mogilner} have only considered cargo-mediated interactions. One of our main purposes here is to study the role of direct motor-motor interaction within such latter modeling framework allowing for non uniform load sharing. In order to do so, we introduce a model similar to one of those proposed in \cite{grossCB2008}, with the addition of direct motor-motor interaction and the allowance of motor back steps. As a counterpart, it considers a less
detailed description of the kinesin cycle, and of the dependence of single motor dynamics on ATP concentration.
The model focus on transport of a cargo by multiple kinesins, but it may be easily adapted to other systems. It considers 
a continuous Langevin dynamics for the cargo while individual motors are assumed to execute discrete jumps
on a one dimensional substrate representing the microtubule. In other related contexts \cite{prlcampas,brugues2009} the role of direct interaction between motors has been analyzed in models which consider the load applied only to the leading motor.

Our work does not focus on the reproduction of particular experimental data but on
analyzing the role of different ingredients that the kind of models considered may
incorporate. The main ingredient is the direct motor-motor interaction, for which we
assume an {\em excluded-volume} type. We analyze both the cases of motor motion on a
single track and on multiple tracks, with the main finding that for realistic numbers of
motors and tracks, motor interactions lead to relevant changes on the load-velocity curve
for cargo when comparing to the non interacting situation. We also study the influence of
motor back steps and of different assumptions for motor detachment.
Other model ingredients whose influence on cargo dynamics has been analyzed in previous
works (although without direct motor-motor interaction), such as the medium viscosity
and the stiffness of the motor-cargo links, are considered here as fixed at standard
experimental values.

The paper is organized as follows. In section 2 we present the model. Section 3 contains
the main analysis of cargo transport by interacting motors moving on a single track,
including the results for the load-velocity curves for cargo and the run lengths. Section
4 studies the case of motion of interacting motors on multiple tracks. In section 5 we
analyze the influence of motor back steps and of different models for detachment. Section
6 is devoted to our conclusion and some final remarks.

\section{The model}

Each motor is modeled as a particle that can occupy discrete positions $x_j=j \Delta x$,
with integer $j$ and $\Delta x=8\, nm$, in a one dimensional substrate representing the
microtubule. Its dynamics is determined by four experimentally accessible quantities,
which are the dwell time $\tau_D$ \cite{cross2005,block2000}, the forward-backward
ratio of jumps $R$ \cite{cross2005,block2000,hyeon09}, the detachment rate $P_{det}$ and the
attaching rate $\Pi$. The first three depend on the load force ($F$) acting on the motor,
while for the attaching rate we consider the fixed value $\Pi=5 s^{-1}$ in agreement with
other studies \cite{lipowskyPhysE,grossCB2008,mogilner}. The dwell time together with the
ratio $R(F)$, both considered as functions of the load force, determine the step
probability per unit time for right (forward) and left (backward) jumps entering in the
Monte Carlo algorithm. They are respectively
$P_r(F)=\left[R(F)/(1+R(F))\right]/\tau_D(F)$ and
$P_l(F)=\left[1/(1+R(F))\right]/\tau_D(F)$. The mean velocity at a fixed load force is
just $v(F)=\Delta x (P_r(F)-P_l(F))$. The details of the Monte Carlo algorithm are given
in the Appendix.

\begin{figure}[h!]
\centering \resizebox{\columnwidth}{!}{\includegraphics{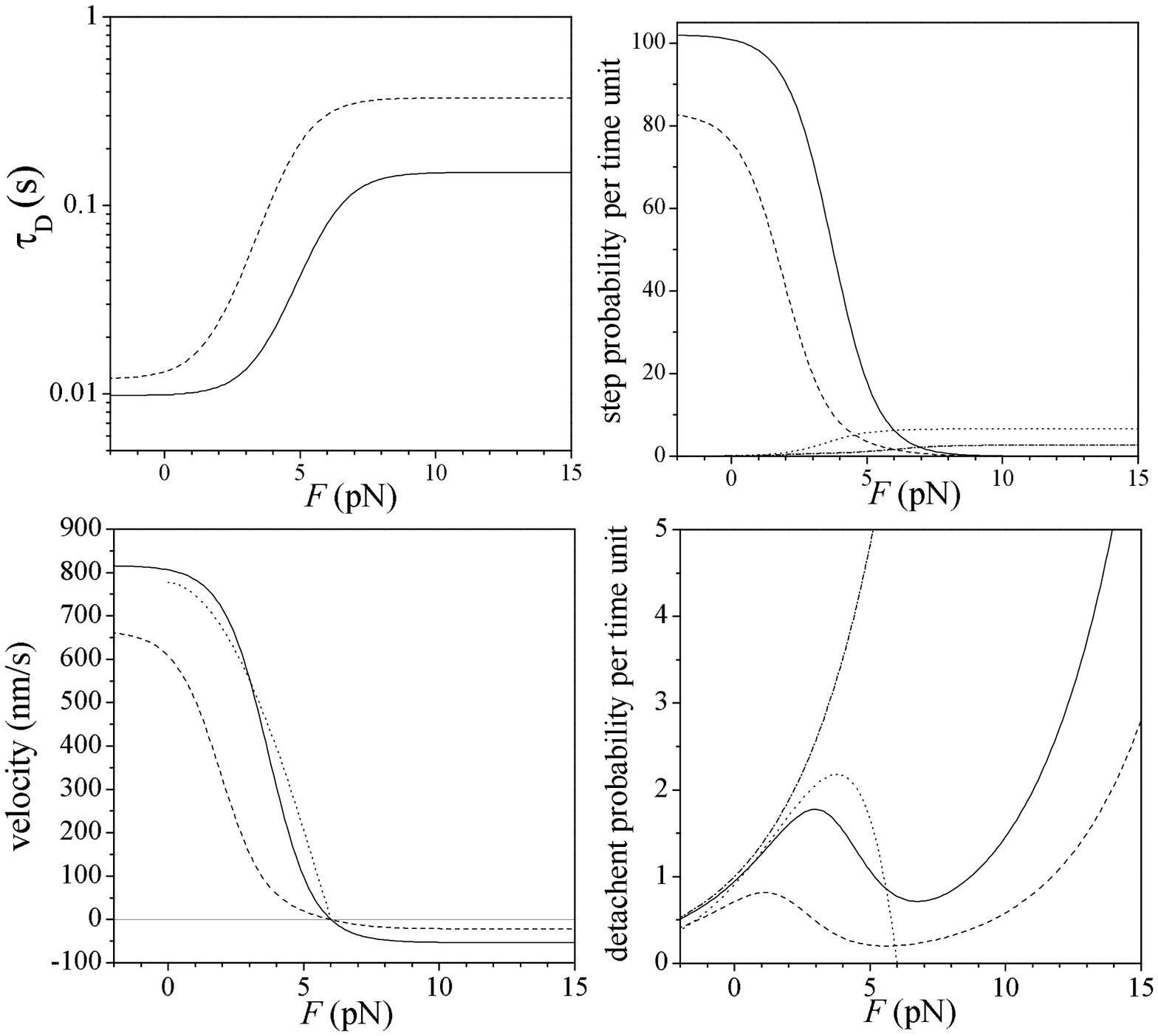}}
\caption{\label{fmodels} Properties defining single motor dynamics as function of the load force. a) Dwell time for
models H (solid line) and L (dashed line). b) Step probabilities for: model H forward (solid line),
model L forward (dashed line), model H backward (doted line), model L backward (dash-dotted line).
c) Motor mean velocity for model H (solid line), model L (dashed line) and model in Ref.\cite{grossCB2008}
for $\left[ATP\right]=1mM$ (dotted line). d) Detachment probabilities: $P_{det}(F)$ for model H (solid line),
$P_{det}(F)$ for model L (dashed line), detachment rate of model in Ref.\cite{grossCB2008}
at $\left[ATP\right]=1mM$ (see text for explanation)(dotted line), and $P_{det}^{exp}(F)$ (dash-dotted line).
The results of model in Ref.\cite{grossCB2008} here shown correspond to our own calculations using the formulas published
in such reference.}
\end{figure}

In our calculations we consider two different $\tau_D(F)$ relations compatible with
observations for kinesin at two different (although not exactly specified) ATP
concentrations \cite{cross2005,block2000}. For both we use the same general form
$\tau_D(F)=a_1+a_2\left[1+\tanh(a_3(F-a_4))\right]$ with two different sets of the
parameters $a_j$, $j=1,...,4$, that are indicated in the Appendix. In this way we obtain
two versions of the single motor model, here referred to as H and L, representing high
and low ATP concentrations respectively. Since the forward backward ratio of jumps is approximately
independent of the ATP concentration \cite{cross2005,gao2006,hyeon09}, we consider the same
function of the load force for both models. Namely: $R(F)=A \exp(-\log(A) F/F_0)$
\cite{cross2005,block2000,gao2006}, with $A=1000 s^{-1}$ and $F_0=6pN$. The parameter
$F_0$ is the stall force for a single motor, which leads to equal left and right step
probability and, consequently, to zero mean velocity. The role of back steps on the dynamics
of a single kinesin motor at the stall force was discussed in reference \cite{cross2006}
and also in \cite{gao2006}. In figure \ref{fmodels}.a we show the $\tau_D$ vs. $F$ curves for
models H and L, while figure \ref{fmodels}.b shows the corresponding right and left step
probability per unit time. In figure \ref{fmodels}.c we show the mean velocity curves for
both models, together with the $v(F)$ curve of the model in \cite{grossCB2008} for ATP
concentration equal to $1mM$ (which is only defined for the range $0\le F\le F_0=6pN$).
As can be seen, model H has a similar $v(F)$ relation to that in
\cite{grossCB2008,block2000} for $[ATP]=1mM$, although with a slightly different
curvature. Something similar occurs with model L and model in \cite{grossCB2008} for
$\left[ATP\right]\simeq 0.18 mM$ and $F_0=6pN$. Hence, such values of ATP concentration
can be considered as reference ones for models H and L. Most of our analysis on the effect of
motor interaction will be performed considering model H, while model L will be used
mainly in order to show the robustness of the general framework.

Following reference \cite{grossCB2008}, in most of our work we consider the detachment
rate as proportional to the step probability. Actually, to the inverse of the dwell time.
We define $P_{det}(F)=\exp{\left(-F/F_d\right)}/(A \tau_D(F))$, with $F_d=3.18 pN$ and
$A=107$\cite{grossCB2008}. Note, however, that our model for detachment differs from that
in \cite{grossCB2008}, since in such work, following the kinesin model in
\cite{block2000}, a more complex description of the kinesin cycle is provided and two
different detaching rates are considered. Our choice is actually based on one of such two
mechanisms (the one acting after ATP binding, which has a larger detachment probability
per time unit). An interesting experimental analysis of the relation between detachment
and ATP binding is given in \cite{toprak2009}. Other recent theoretical works, such as
\cite{mogilner} and \cite{lipowsky09} considers the pure exponential form
$P_{det}^{exp}(F)=\epsilon_0 \exp{(-F/F_d)}$ with $\epsilon_0\simeq 1 s^{-1}$ and
$F_d\sim 3 pN$. In section 5 we discuss the influence of different models for detachment
on the dynamics of cargo, including $P_{det}^{exp}(F)$. Nevertheless, the main
conclusions of our work concerning the effects of interactions are independent of the
specific prescription for the detachment rate. In figure \ref{fmodels}.d we show the detachment
rates $P_{det}(F)$ for model H and L, together with our calculation of the {\em after-ATP} detachment rate
of the model in reference \cite{grossCB2008} for $\left[ATP\right]=1mM$, and the
exponential formula $P_{det}^{exp}(F)$ for $F_d=3.18 pN$. Recall, however, that the
detachment mechanism considered in reference \cite{grossCB2008} does not act on the whole
kinesin cycle but only on a particular stage \cite{grossCB2008}. So, the comparison of
the corresponding formula with our $P_{det}(F)$ and with $P_{det}^{exp}(F)$ should only
be considered qualitatively.

In order to model the transport of a cargo by multiple motors we follow an approach
similar to those in \cite{grossCB2008} and \cite{mogilner}. The cargo is modeled as a
particle that performs a continuous overdamped Brownian motion in one dimension,
influenced by the forces coming from each of the $N$ motors considered, plus an external
(load) force and thermal noise \cite{grossCB2008}. The cargo is linked to each motor by a
non linear spring \cite{grossCB2008,mogilner} which produces only attractive forces, and
only for distances larger than a critical one. Let us call $x_c$ the position of the
cargo, $x_i$ ($i=1,...N$) the (discrete) positions of the motors and $\Delta_i=x_i-x_c$.
Each motor exerts a force $f_i$ on the cargo, which is defined as $f_i=k(\Delta_i-x_0)$
for $\Delta_i\ge x_0$, $f_i=0$ for $-x_0<\Delta_i< x_0$ and $f_i=k(\Delta_i+x_0)$ for
$\Delta_i\le -x_0$, with $x_0=110\, nm$ and the elastic constant $k=0.32\, pN/nm$
\cite{grossCB2008,mogilner}. Note that the consideration of the cargo as a point particle 
implicitly include the assumption that all the motors are attached to a single spot on the cargo, 
in agreement with the considerations in \cite{grossCB2008} and experimental evidence in \cite{grossCB2007}. 
Even though this is not the most general hypothesis as in vitro experiments indicate \cite{grossPNAS2007,Beeg08},
we will limit our analysis to such assumption. The dynamical equation considered for the cargo is
\begin{equation}
\label{cargodyn}
\gamma \frac{d x_c}{d t}=-L+\xi(t)+\sum_{i=1} f_i,
\end{equation}
where $\gamma$ is the viscous drag, $L$ the external load force and $\xi(t)$ the thermal
noise. We consider the fixed value $\gamma=9.42\,10^{-4} pN s/nm$ defined through the
stokes formula \cite{grossCB2008,mogilner,grossCB2007} using a viscosity equal to $100$
times that of water and a radius of the cargo equal to $0.5 \mu m$. The dynamics
of the $i$-th motor is ruled by the Monte Carlo algorithm with the instantaneous load
force $F=f_i$. Just after a motor detaches from the microtubule, and during the time it
remains detached, its position can be considered as equal to that of the cargo (i.e. the
motor does not support force). The attachment of detached motors occurs with equal
probability in any of the discrete sites $x_j$ satisfying $|x_j-x_c|<x_0$.

The interaction between motors is modeled by including in the Monte Carlo algorithm a
constraint that forbids two motors to be at the same site. This means that a motor can
only perform a forward (backward) step if the right (left) site is empty. See algorithm 
details in the Appendix.

In order to identify the effect of motor interactions on the dynamics of cargo,
we will compare the results of our model of interacting motors (IM) with results for non
interacting motors (NIM). For the sake of completeness, we will also consider the IM
model without detachment.

The main relevant quantities to compute using the different models are the cargo mean
velocity as function of the load force $v(L)$, the run length $r(L)$, defined as the mean
distance traveled by the cargo before all the motors become detached, and the stall
force for cargo $L_0$, which is the value of $L$ at which $v(L)$ vanishes. No
confusion should occur between $L_0$ and the stall force for a single motor $F_0$, which
is a system parameter. Other quantities of interest are the mean force acting on the
motors, the mean number of attached motors at a given time, and the mean number of
pulling motors. Precise definitions for such quantities will be given in the following
sections.

Numerical simulations are performed using the algorithms explained in the appendix. For
$v(L)$ calculations we typically consider run times of $10s$ with $dt=2 \times10^{-5}s$, and we 
average over a number of realizations which ranges between $2000$ for $N=2$ and $100$ for $N=12$.
Results for run times are obtained from long time realizations which last until complete
detachment. As initial conditions we consider that all the motors are attached at random positions.

\section{Interacting motors on a single track}

\subsection{Cargo velocity vs. load force.}

In this subsection we present the load-velocity curves for IM and NIM models considering
both H and L dynamics for single motors. We here limit our presentation to
describe and compare the results from the different models, while most of the
interpretation of the results will be given in the next subsection on the basis of the
analysis of the spatial distributions of motors and force sharing behavior.

\begin{figure}[h!]
\centering \resizebox{\columnwidth}{!}{\includegraphics{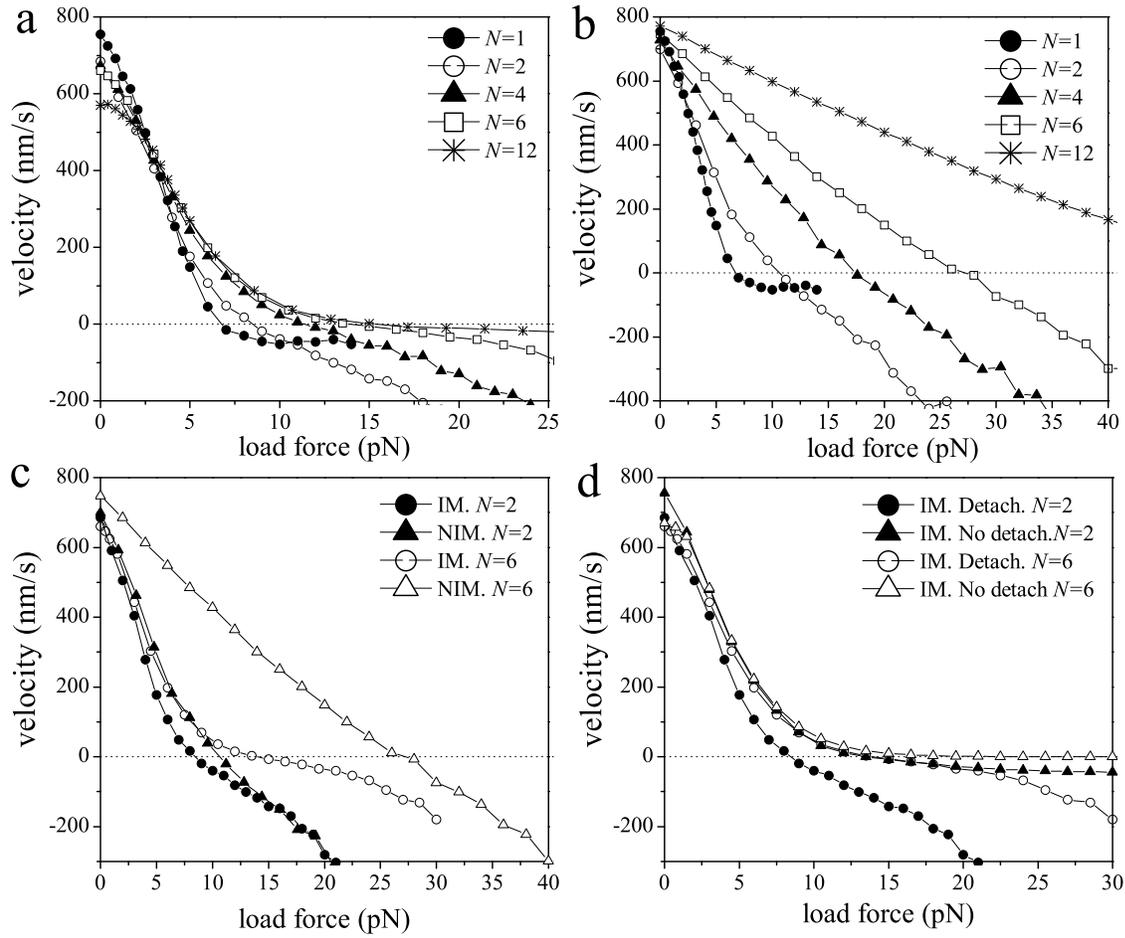}}
\caption{\label{fMHv248D} Cargo mean velocity vs. load force for model H. a) Results for IM considering different values of $N$. b) Ibid (a) for NIM. c) Comparison of the results for IM and NIM considering $N=2$ and $N=6$. d) Results for
IM with and without detachment for $N=2$ and $N=6$.}
\end{figure}

In figure \ref{fMHv248D} we show the main results for $v(L)$. The different panels allow
us to analyze the dependence on $N$ and the effects of motor interactions and motor
detachment as well. In all calculations, individual motor dynamics is ruled by model H.

Firstly, it can be seen that within all the models $v(L)$ decreases monotonously with
$L$. In fact, except for the case of interacting non detaching motors at large $N$
(panel d), $v(L)$ become negative for large enough $L$, meaning that the cargo moves
effectively backward. As we will see later, the corresponding run lengths for such
backward motions are, in most cases, non negligible. The possibility of observing 
sustained back motion at large  $L$ is in accordance with the possibility of a {\em tug
of war} dynamics \cite{tugofwar}, if it is assumed that the external force is due to 
a team of motors of a different class which advances in the opposite direction. Note that
in figures \ref{fMHv248D}.a and \ref{fMHv248D}.b the results for $N=1$ are shown only for $L\lesssim 15pN$.
Beyond such value of $L$ the corresponding run lengths become very small and
the motion of cargo can hardly be considered as processive.

The difference between the $v(L)$ curves for IM and NIM models with detachment is
apparent. Figures \ref{fMHv248D}.a and \ref{fMHv248D}.b allow us to compare the 
dependences on $N$. Leaving aside the small $L$ behavior,
which deserves a special treatment that will be given in Section 3.3, we see that
$v(L)$ increases with $N$ at fixed $L$, both for IM and NIM. However, the increasing is
different for each model. For IM we have that the growing of $v(L)$ saturates to constant
values at relatively small $N$. In fact, except for very large load forces leading to
back motion of cargo, the saturation occurs for $N\gtrsim 6$. Note that, in particular,
the stall force $L_0$ remains essentially independent of $N$ for $N\gtrsim 6$. As we will
show in the next subsection, this saturation effect is due to jamming phenomena which
limit the number of motors that share the force. 
For NIM, in contrast, no jamming effects are present and the increasing of $v(L)$ with $N$
seems to be limited only by the free kinesin velocity ($v\sim 800 nm/s$), which could be
eventually attained at high enough $N$. The saturation observed for IM was also predicted
in \cite{prlcampas}, where a simpler and less detailed model for interacting motors was
studied. This coincidence stress the relevance of such kind of minimal but thorough
models capturing the essential features involved in the observed phenomena.

Figure \ref{fMHv248D}.c shows that, for fixed $N$, IM lead to lower cargo velocity and
stall force than NIM. Moreover we see that the differences between the curves for both
models increase with $N$ but vanish at very large $L$, when fast back motion of cargo is
observed.

Concerning the comparison of the models of IM with and without detachment, the results in
figure \ref{fMHv248D}.d show that detachment leads to lower values of $v(L)$ and
$L_0$. However, at large $N$, the differences between both models become relevant
only for large $L$, while for relatively small load forces both models give very similar 
results. Note that the allowance of
detachment turns out to be relevant for increasing the magnitude of the negative velocity at
large $L$. In fact, in the following sections we will show that detaching is much more
important than motor back steps in what concerns sustaining back motion of cargo.

\begin{figure}[h!]
\centering \resizebox{\columnwidth}{!}{\includegraphics{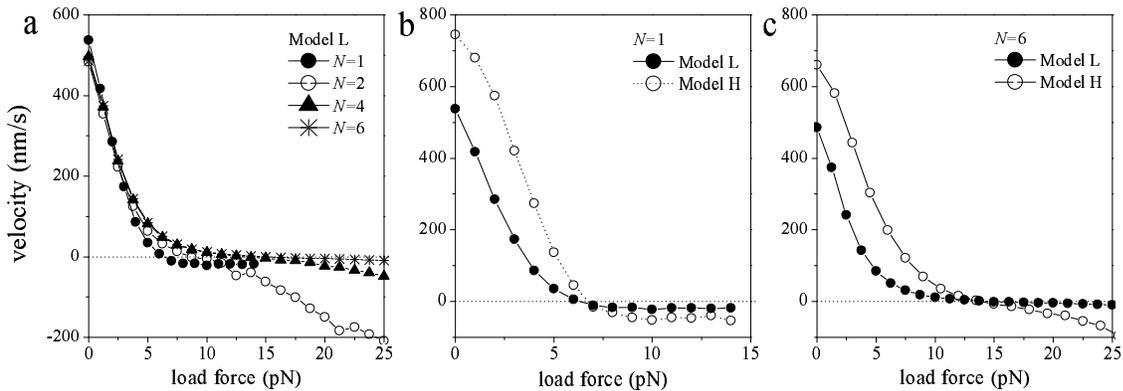}}
\caption{\label{fMHMLv} Mean cargo velocity vs. load force. a) Results for model L for different values of $N$.
b) Comparison of models L and H with $N=1$. c) Ibid (b) for $N=6$.}
\end{figure}

Completely analogous results to those shown in figure \ref{fMHv248D} were obtained by
considering model L instead of model H, with the same conclusions on the role of
motor interaction and detaching. In figure \ref{fMHMLv}.a we show the force velocity
curves for IM using model L for varying $N$. In figures \ref{fMHMLv}.b and \ref{fMHMLv}.c
we compare the $v(L)$ curves for models L and H for $N=1$ and $N=6$. The results
show the way in which cargo dynamics is affected by a change in the ATP concentration.
Note that model H produces higher absolute velocities for both forward and backward
motion of the cargo.

\begin{figure}[h!]
\centering
\resizebox{\columnwidth}{!}{\includegraphics{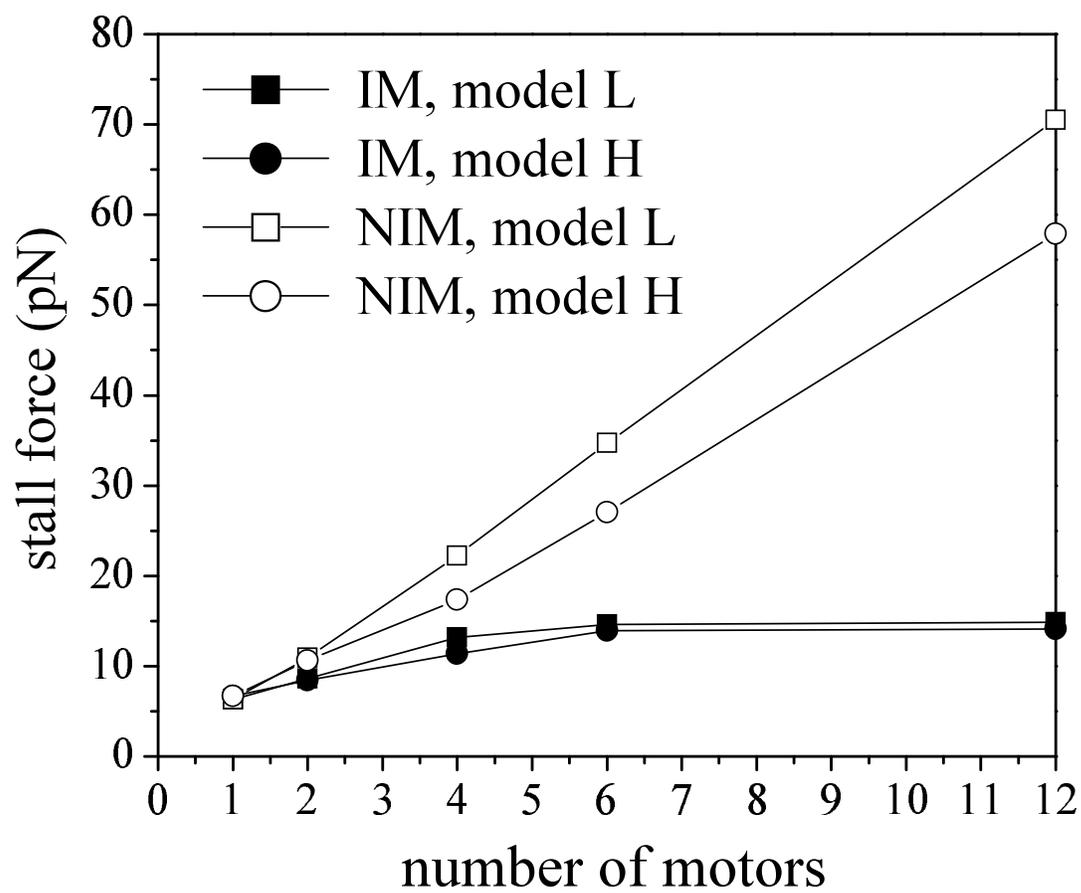}}
\caption{\label{stallvl0} Stall force as functions of the number of motors for models H and L. Results for IM and NIM motors
with allowed detachment.}
\end{figure}

Finally, in figure \ref{stallvl0} we show the stall force $L_0$ as function of $N$ for
NIM and IM models considering both L and H single motor dynamics. We find that the stall
force grows in a quasi linear way for NIM, while for IM it shows the mentioned
saturation effect.

\subsection{Motor distributions, trajectories and force sharing.}

The results for the velocity of cargo as a function of $L$ presented above can be
understood by analyzing the dynamics of individual motors linked to the cargo and the
forces acting on them. The detailed way in which the motors are distributed in space provides
relevant information on these items and on the kind of dynamics that could be expected.
We thus analyze the spatial distribution of motors relative to the cargo position in the
long time regime, here referred to as $P(x-x_c)$. We compute it form histograms of
the positions of attached motors and we normalize it in such a way that
$\int_{-\infty}^\infty P(x-x_c) dx$ is equal to the asymptotic mean fraction of attached
motors (i.e. the time average of the number of attached motors divided by $N$). Thus, $N
P(x-x_c)$ represents the probability density of finding a motor at $x$ given that the
cargo is at $x_c$. Recall that the natural spring length is $x_0=110nm$. Only the motors
which are beyond such limit exert force on the cargo and, in turn, their dynamics are affected 
by the corresponding tensions on the springs. We refer to such motors as the {\em pulling
motors}. The mean number of pulling motors at a given time, given by $\int_{x_c+x_0}^\infty N P(x-x_c) dx$, 
is a relevant quantity for understanding the behavior of $v(L)$. 

\begin{figure}[h!]
\centering
\resizebox{\columnwidth}{!}{\includegraphics{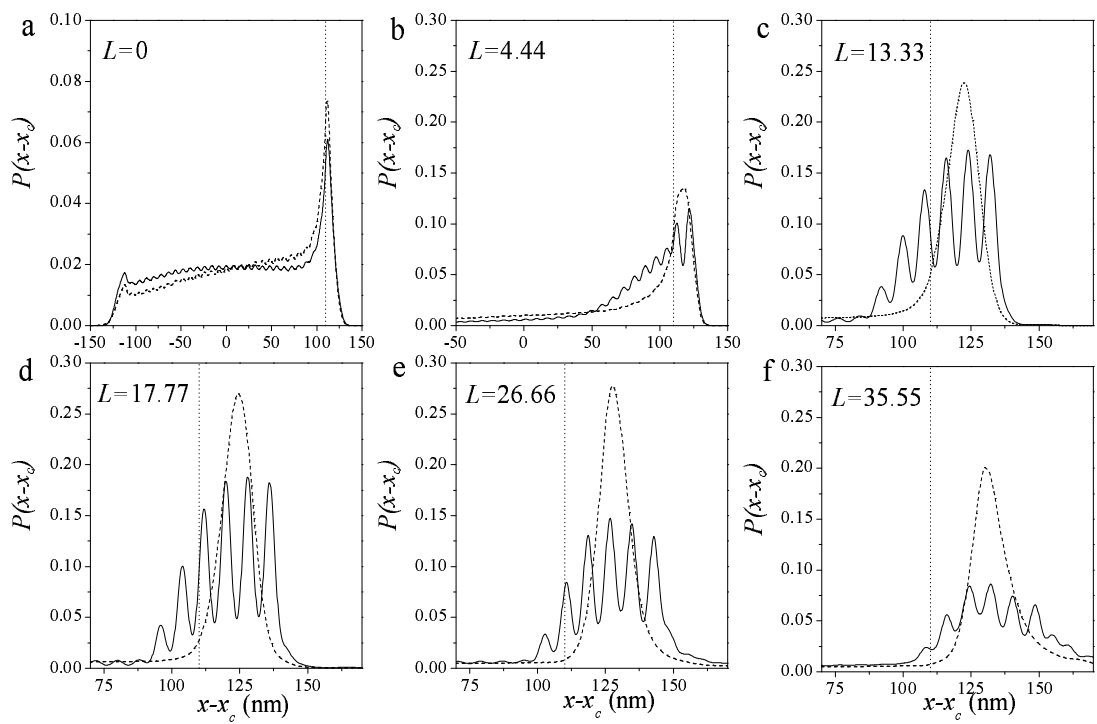}} \caption{\label{fdensity} Spatial distribution 
of motors as function of the distance to the cargo. Results for IM
and NIM considering model H with $N=6$ and different values of load force. In all panels
the vertical dotted segment indicate distance $x_0$ beyond which motors exert non
vanishing forces on cargo.}
\end{figure}

In figure \ref{fdensity} we show the distribution $P(x-x_c)$ for $N=6$ at different
values of $L$ considering IM and NIM models. Motors tend to accumulate to the right of
the cargo, at distances close to $x_0$. This effect,
which occurs for both IM and NIM, is due to the fact that free motors (those at
$|x-x_c|<x_0$) advance faster than cargo. The probability of finding a motor at
$(x-x_c)<0$ is appreciable only for very small $L$ (see panel (a) and
discussion on the next subsection), for which the difference between cargo and free
motors velocities is relatively small. Clearly, such difference increases with $L$ as
cargo velocity decreases. Since motor interaction forbids two motors to occupy the same
site, IM lead to wider distributions than NIM, and with multiple peaks. The peaks are separated 
according to the periodicity of the substrate by distances equal to $\Delta x= 8 nm$.

Let us first analyze figures \ref{fdensity} a and b for which we have
forward motion of cargo for both models. Clearly, there is a maximum distance from the
cargo that motors attain which is equal for IM and NIM. It is approximately
$(x-x_c)\sim134nm$. Such maximum distance can be understood as a consequence of the
competition between advance and detachment occurring at $(x-x_c)>x_0$. As a motor moves
away from the cargo to distances much larger than $x_0$, its detaching
probability increases exponentially with the distance,  while its forward step probability
decreases rapidly to zero and its back step probability increase to a maximum asymptotic
value. These dependences are shown in figure \ref{fdetachdex}. It can be seen that
$P_{det}$ equals $P_r$ at $(x-x_c)\simeq134.5nm$. Hence, for $(x-x_c)>134.5nm$ we have
$P_r<P_{det}$, so a motor will unlikely advance before detaching. Moreover, in the range
$130nm<(x-x_c)<155nm$ the back step probability $P_l$ is higher than both $P_r$ and
$P_{det}$. Thus, a motor in the domain $130nm<(x-x_c)<134nm$ will most probably jump
back. These observations explain the peak at $\sim 120nm$ of the distribution for NIM. In
the case of IM, the maximum distance from cargo is the same as for NIM, but the
distribution of motors spread to the left due to the effective repulsion between motors.
Due to this, the number of pulling motors is sensibly lower for IM than for NIM. This is
the main cause of the lower values of cargo velocity observed for the IM model for
forward motion of the cargo.

Note that the peaks of the distribution for IM become more notable as $L$ is increased
from $L=0$ to $L=17.77pN$. This is because, the amplitude of the peaks are determined by
fluctuations of the position of the cargo,
and such fluctuations are reduced when increasing the tension applied on the cargo by the
simultaneous action of the force $L$ to the left and the spring forces to the right.

The results in figure \ref{fdensity}.c correspond to a load force equal to the stall
force for IM. Figures \ref{fdensity}.d, e and f, show the distributions for
larger values of $L$ for which IM produce back motion of cargo, while NIM lead to forward
(panels d and e), or back (panel f) motion of the cargo. In panels d and e we see
that $P(x-x_c)$ has non vanishing contributions at larger distances for IM than for NIM.
This is mainly due to that, for back motion of the cargo, larger values of $(x-x_c)$ can
be attained not because of the advance of the motors, but because of the back motion of the
cargo. As we will later see when studying the motors trajectories, the more likely
mechanism is such that the last forward motor step occurs when $(x-x_c)\lesssim 130nm$ but
then the cargo continues moving back and the distance increases.

When changing from panel d to f we see that the area below the distributions decreases
for both models. This is because the fraction of attached motors decreases due to the
increasing of the load force and the detaching probabilities. For very large $L$
as that on panel f, the peaks of the distribution for IM tend to disappear. This is
due to the relatively high frequency of detachments, which makes the cargo position
to fluctuate (continuously) through several $8nm$ periods while attached motors remain
stall.

\begin{figure}[h!]
\centering
\resizebox{\columnwidth}{!}{\includegraphics{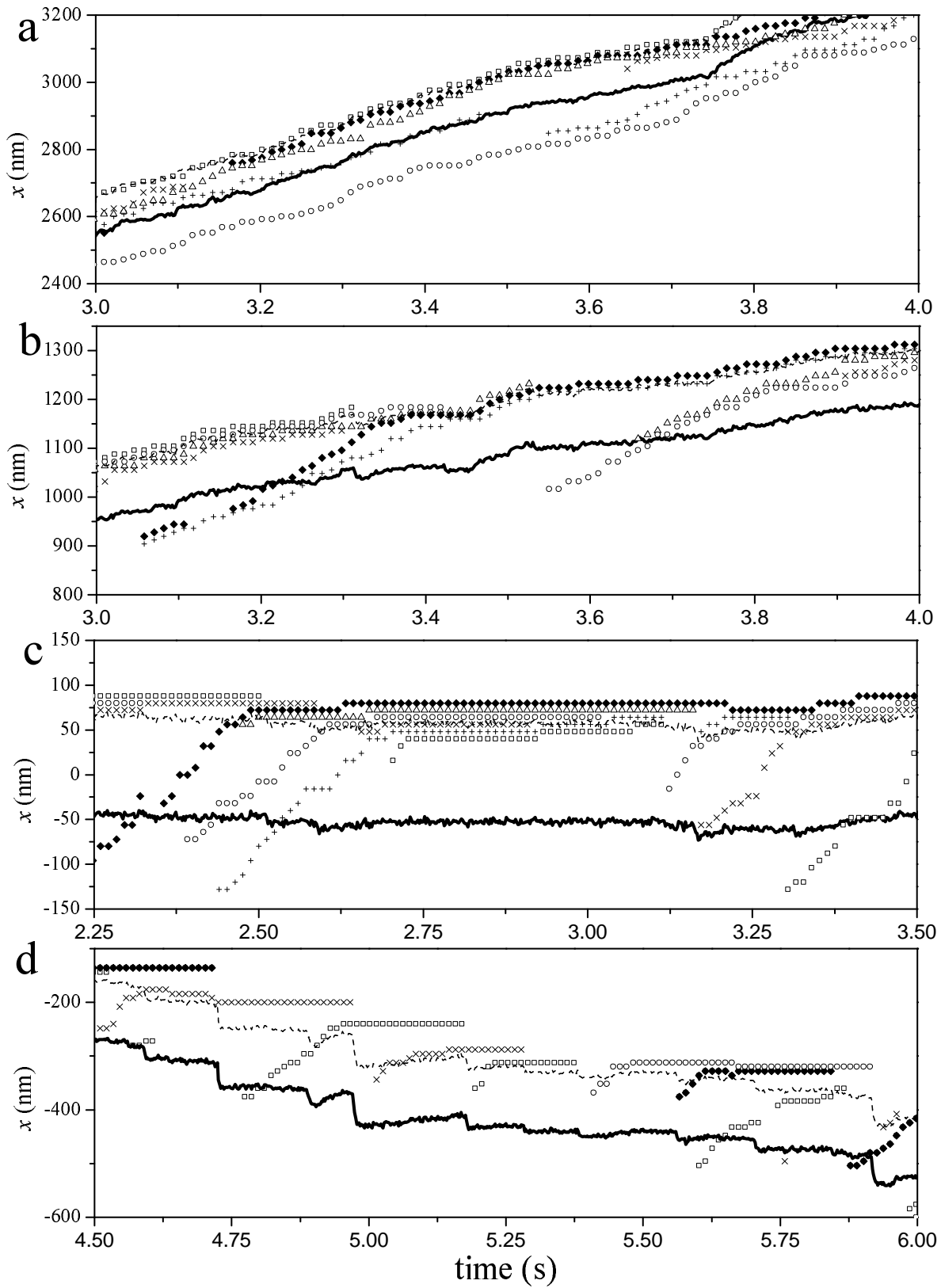}}\caption{\label{trajec} Trajectories of
motors (symbols) and cargo (thick solid line) for a system of 6 motors using
model H. Results for: (a) $L=0$, (b) $L=4.44$, (c) $L=13.33$ (stall force) and $L=26.66$.
In the four panels the dashed line indicates a distance from cargo equal to $x_0=110 nm$.}
\end{figure}

In figure \ref{trajec} we show trajectories of the cargo and motors for $N=6$ considering
IM ruled by model H. Trajectories in panels a, b, c and d, correspond to the motor distributions (and
system parameters) in figures \ref{fdensity} a, b, c and f respectively. In all panels
the dashed line indicates a distance $x_0$ from cargo. The motors that are above such
curve are the pulling motors. Let us first analyze figure \ref{trajec}.a which
corresponds to $L=0$. In this case, the cargo velocity is almost equal to the free motors
velocity. Thus, some motors may advance behind cargo during relatively long periods of
time. In figure \ref{trajec}.b the cargo advances slower than free motors and motors
accumulate near $(x-x_c)\sim x_0$ (see motor distribution in figure \ref{fdensity}.b).
Due to the interactions, the motors organize in a queue that advance in the preferred
direction ahead of the cargo. For the case in figure \ref{trajec}.b there are typically two 
pulling motors at a given time, corresponding to the two peaks of the $P(x-x_c)$ that are beyond
$(x-x_c)=110 nm$ in figure \ref{fdensity}.b. Immediately below the pulling limit, other
one or two motors are waiting for their turns for pulling, that will come when one of the
pulling motor detaches. The velocity of such {\em waiting} motors is limited by the
interaction with forward motors and not by the forces coming from the cargo which are null.
At lower values of $(x-x_c)$ it is possible to observe, from time to time, the recently
attached motors that advance freely until they arrive to the rear part of the queue. Such
motors contribute to the tail of the $P(x-x_c)$ distribution at low $(x-x_c)$. These results
are compatible with the observation that, at small load, some motors do not contribute
constructively to cargo velocity but they do to processivity
\cite{grossCB2007,rogers2009}.

Figure \ref{trajec}.c shows the trajectories of cargo and motors for $L=L_0$. Clearly, the
mean velocity of the cargo is zero, while fluctuations of the cargo's position are of the
order of $20nm$. The queue dynamics of the motors is similar to that in figure \ref{trajec}.b.

Figure \ref{trajec}.d shows the trajectories for a large value of $L$ for which we observe a 
rapid backward motion of cargo. Now, the motors pull in opposition to the advance of the cargo, 
so they act effectively as breaks. The number of pulling motors is larger than for $L\le L_0$. 
This is due to the effect explained before connected to the back motion of the cargo. Note that 
the advance of motors occurs for $(x-x_c)<130nm$ and, then, the distance increases due to the 
back motion of the cargo.

Another interesting result is that backward motion of cargo occurs at intermittent periods associated with detaching of motors, while the rest of the time a zero velocity regime is observed. Thus, motor detachment appears as the main dynamical mechanism responsible for back motion of cargo. Motor back steps has a secondary role which, as we will discuss in section 5, is only slightly relevant at smaller values of $L$ and also for models of NIM in more general conditions.

\begin{figure}[h!]
\centering \resizebox{\columnwidth}{!}{\includegraphics{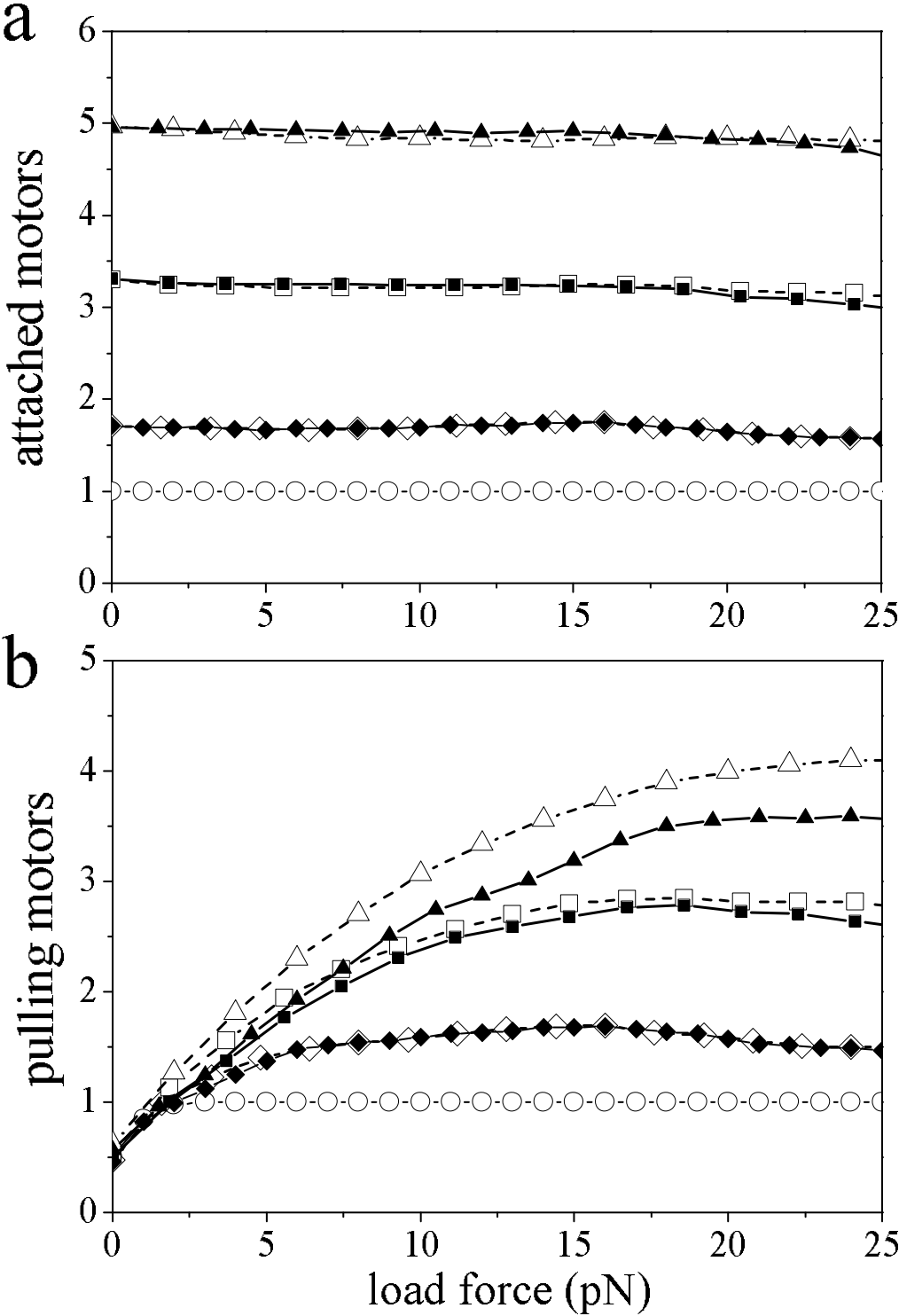}} \caption{\label{fpulling} a) Mean
number of attached motors. Results for $N=1$ (circles), $N=2$ (rhombus), $N=4$ (squares) and $N=6$ (triangles).
Solid symbols correspond to IM while open symbols indicate results for NIM. b) Ibid (a) for the mean number of
pulling motors.}
\end{figure}

In figure \ref{fpulling} we show the mean number of attached motors and the mean number
of pulling motors as functions of $L$ for different values of $N$. The
number of attached motors results quite similar for IM and NIM models for all values of $L$ and
$N$ analyzed. In contrast, the number of pulling motors is larger for NIM than for IM, except 
for very small values of $N$ and $L$. The differences are clearly observable for $N>2$ and $L>1pN$. 
The behavior at small $L$ will be discussed later in more detail.

Recall that the larger number of pulling motors observed for NIM was indicated
as the main cause of the cargo velocities being larger for NIM than for IM. This is true for $N>2$, but,
in the case $N=2$, for which the number of pulling motors is equal in both models, a different argument should
be given. The explanation will come through the analysis of the mean forces acting on the different motors.
In order to compute them we relabel the motors from $1$ to $N$ at each time, in such a way that
motor $1$ is the one that goes ahead, motor $2$ is the one that goes second, and so on. The detached motors are
considered at the end of the list. For instance, in case that they were two detached motors, they would be
indistinctly labeled as $N-1$ and $N$. In the NIM model, for which two or more motors can be
at the same site, the sub-order of the motors at the same position is randomly
assigned. (For instance, if two motors share the leading position, they are indistinctly
labeled as $1$ and $2$.) We then compute the forces on the relabeled motors getting
$f_1\ge f_2\ge f_3 ...\ge f_N$. Finally we perform time averaging to obtain the mean
force acting on the $j$-est motor $\langle f_j\rangle$. Note that the sum $\sum_j \langle
f_j \rangle$ is not expected to coincide with $L$ but with $\gamma \langle dx/dt\rangle+L$ 
(see equation (\ref{cargodyn})).

\begin{figure}[h!]
\centering \resizebox{\columnwidth}{!}{\includegraphics{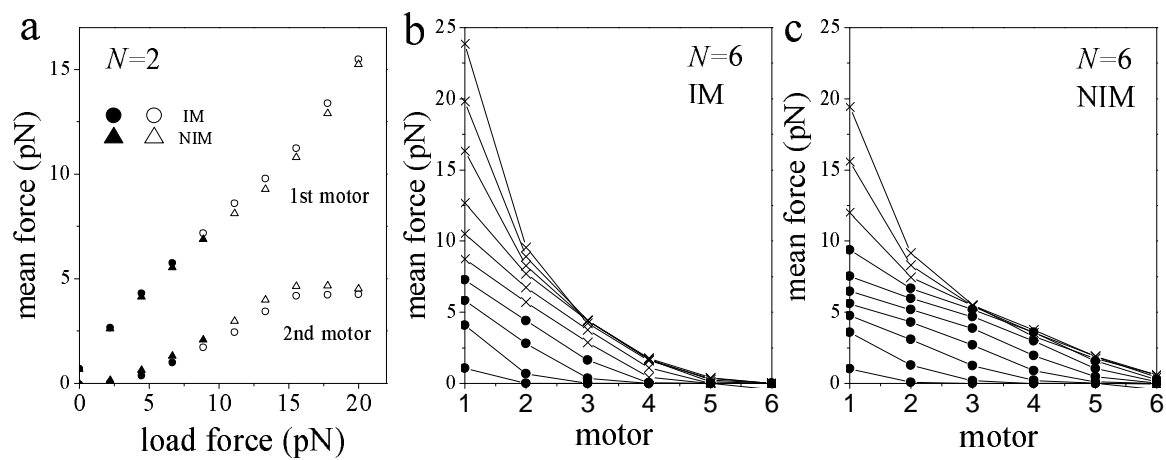}}
\caption{\label{fuerzas} Mean forces acting on the different motors for IM and NIM using model H. a) Results for $N=2$
for the mean force on the first and second motor as function of $L$. Circles correspond to
IM while triangles indicate results for NIM. Solid and open symbols correspond to forward and backward
motion of cargo respectively. b) Results for $N=6$ for IM. Each curve is for a different value of $L$
ranging equidistantly from $L=0$ (bottom curve) to $L=40$ (top curve). Solid circles indicate forward motion
of cargo while crosses indicate back motion of cargo. c) Ibid (b) for NIM.}
\end{figure}

In figure \ref{fuerzas} we show mean-force results comparing IM and NIM models.
Figure \ref{fuerzas}.a shows the $N=2$ case. The mean forces acting on the first and second motor
are plotted as functions of $L$. It can be seen that the force is better shared between the two motors in
the case of NIM, i.e. the force on the two motors are less different each other for NIM than for IM.
Accordingly, the force on the first motor is larger in the case of IM. This fact clearly explains the
larger velocity found for NIM, since a larger force on the leading motor will produce slower velocities
of the leading motor and cargo. Note that although the differences in the forces and also those in the 
velocities can be considered relatively small, they imply a decreasing of about $20\%$ in the 
stall force when passing from NIM to IM. In figures \ref{fuerzas}.b and \ref{fuerzas}.c we show the mean forces acting 
on the different motors for $N=6$, considering IM and NIM respectively. The curves for 
different values of $L$ reveal that forces are better shared in the case of NIM and that the force on the 
leading motor is larger for IM. This is in agreement with what is expected from the analysis of the 
results for velocities and spatial distributions of motors.

\subsection{Behavior at vanishing and small loads}

\begin{figure}[h!]
\centering \resizebox{\columnwidth}{!}{\includegraphics{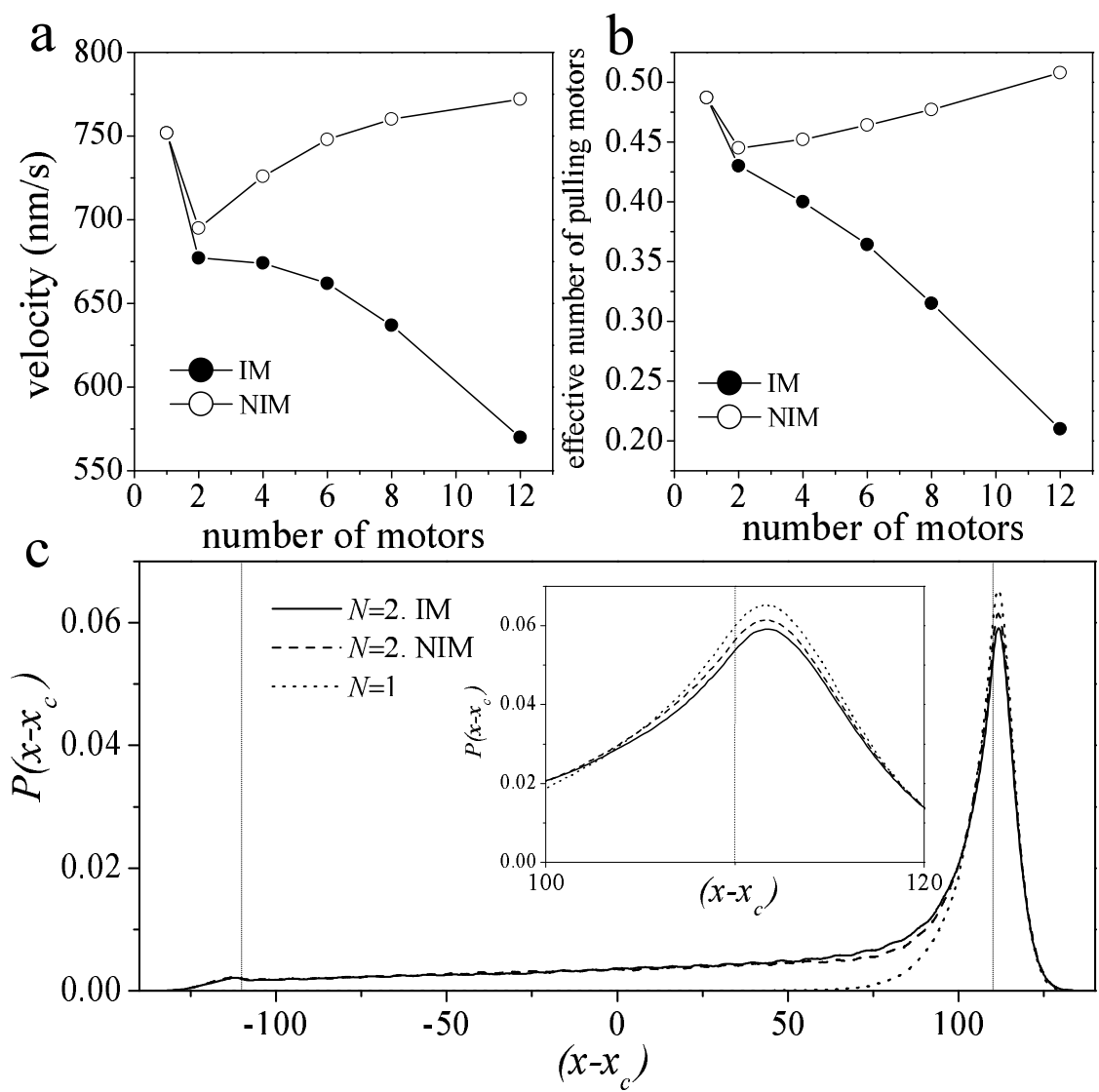}} \caption{\label{fzeroload}
Results at zero load. a)
Cargo mean velocity as a function of $N$ for NIM and IM using model H. b) Effective number of pulling motors.
c) Motor distributions for $N=1$ and $N=2$ for IM and NIM. The vertical segments at $(x-x_c)=\pm110nm$ indicate the
limit positions for pulling. The inset in panel (c) show the details of the distributions
close to the maxima at ($x-x_c\sim112nm$).}
\end{figure}

As stated before, cargo dynamics at small $L$ has some peculiarities. 
On one hand, for the case of IM, figure \ref{fMHv248D}.a shows
us that for $L\lesssim 2.5pN$, $v(L)$ decreases with $N$ instead of increasing as it
happens for larger values of $L$. On the other hand, for NIM, figure \ref{fMHv248D}.b shows
that, at small enough $L$, $v(L)$ depends on $N$ in a non monotonous fashion. These
dependencies are better shown in figure \ref{fzeroload} considering the case $L=0$.
The relevant characteristic of motor dynamics that will allow us to understand these
behaviors is that, at $L\sim0$, the velocity of free
motors (those at $|x-x_c|<x_0$) is quite similar to the cargo
velocity. This causes particularly broad motor distributions $P(x-x_c)$, with
relevant contributions even at $(x-x_c)<-x_0$, as can be seen in the
distributions for $N=2$ shown in figure \ref{fzeroload}.c, and also in the distribution
for $N=6$ in figure \ref{fdensity}.a. and in the trajectory in figure \ref{trajec}.a.
This means that it is possible to find motors that are effectively pulling in opposition
to the advance of the cargo. So, the relevant quantity for understanding the $L\sim0$ behavior is
not the number of pulling motors but what we will call the {\em effective number of pulling
motors}, defined as the difference between the mean number of motors that pull the cargo
forward, and those that pull against the advance of the cargo. This is, 
$\int_{x_c+x_0}^{\infty} N\, P(x-x_c)\, dx-\int_{-\infty}^{x_c-x_0}N\, P(x-x_c)\, dx$. For the
distribution in figure \ref{fzeroload}.c, it corresponds to the difference between the
area at the right of the segment at $(x-x_c)=110nm$ and the area at the left of the segment
at $(x-x_c)=-110nm$. In figure \ref{fzeroload}.b we plot the effective number of pulling
motors as function of $N$ for IM and NIM. The correlation between such results and those
for the velocities shown in figure \ref{fzeroload}.a is apparent.

Regardless the differences between the cases of IM and NIM models, the fast drops
occurring both in the velocity and in the effective number of pulling motors when passing
from $N=1$ to $N=2$ can be understood by observing that, for $N=1$, the $P(x-x_c)$
distribution has null contributions at negative values. Moreover, in most of the region
where it is non vanishing, it coincides quite well with the distribution for $N=2$, while
the latter distribution has a much broader tail to the left. In fact, the areas under the
$N=1$ and $N=2$ distributions are equal to $1$ and $1.7$ respectively, which are the
corresponding number of attached motors for each case. Roughly speaking this means that
passing from $N=1$ to $N=2$ corresponds to effectively adding to the system one motor that
never push forward but sometimes pull back. In addition, one of the motors is 
detached part of the time, so that at a given time we have $1.7$ motors on average. Note that, most
of the time, the second motor will neither push forward nor pull back, but will
constitute a kind of spare motor in case the first of the queue detaches. This is the
origin of the enhancement of the run length observed when passing from $N=1$ to $N=2$,
that we will analyze in the next subsection.

The difficulties for motor advance induced by motor interaction affect both the entering
to the $(x-x_c)>x_0$ zone and the exiting from the $(x-x_c)<-x_0$ zone as well. Thus, for large
enough $N$, in the same way that interactions cause a decreasing of the number of pulling
motors, they cause an {\em increasing} of the number of motors that pull against the
advance of cargo. Since interaction and jamming effects increase with $N$, we have that
the effective number of pulling motors decreases with $N$ for the IM model. In contrast,
for NIM, motors advance freely and although the number of pulling-against motors may
increase with $N$, the number of pulling-forward motors increases in a much faster way.
This is the origin of the differences between the curves for the effective number of
pulling motors shown in figure \ref{fzeroload}.b., and ultimately, the explanation for
the behaviors of the velocity curves in figure \ref{fzeroload}.a. Note that in the
distributions for $N=2$ shown in figure \ref{fzeroload}.c, there are no differences
between the values for IM and NIM at $(x-x_c)<0$. This is because the probability of
finding the two motors interacting at $(x-x_c)<0$ vanishes. Interactions occur only in
the region $(x-x_c)\sim x_0$, affecting the number of pulling-forward motors. In contrast,
the distributions shown in figure \ref{fdensity}.a. differ for IM and NIM in the whole
range of $(x-x_c)$. This is because for $N=6$ interactions are possible everywhere, 
affecting also the number of pulling-back motors.

For the case of NIM, the small load behavior described here is completely compatible with
that previously found in \cite{grossCB2008}. The fact that two motors lead to slower
cargo velocity than one motor was explained in such reference as due to the occurrence of rapid back
motion of cargo associated to detachment of the leading motor. This is in complete
agreement with our explanation, since such rapid back motion of cargo would only occur if
the second motor is a pulling-back motor. Moreover, the presence of a pulling-back motor clearly
increases the probability of detachment of the leading motor.

Recent in vivo experiments in Drosophila embryos \cite{cellgross2008} have shown a decrease 
of the cargo velocity when increasing the number of motors. The coincidence of this rather 
counterintuitive find with our results and with those in \cite{grossCB2008} is interesting. 
However, the relation should be taken with care, since experimental results in \cite{cellgross2008} 
show different stall forces to that on our modeling assumptions, and also no dependence of the 
run lengths on $N$, in contrast to what is observed in vitro \cite{grossPNAS2007} and in our 
results (see next Subsection). This last fact may be due to additional factors involved in vivo, 
including higher level mechanisms of run length control \cite{cellgross2008}.

\subsection{Run lengths}

To complement our studies on the cargo dynamics we present here results
for the run length $r(L)$ defined at the end of section 2. Such quantity is easily measurable 
in experiments and of great theoretical relevance \cite{grossCB2008,lipowskiBioPJ,grossCB2007} .

\begin{figure}[h!]
\centering
\resizebox{.8\columnwidth}{!}{\includegraphics{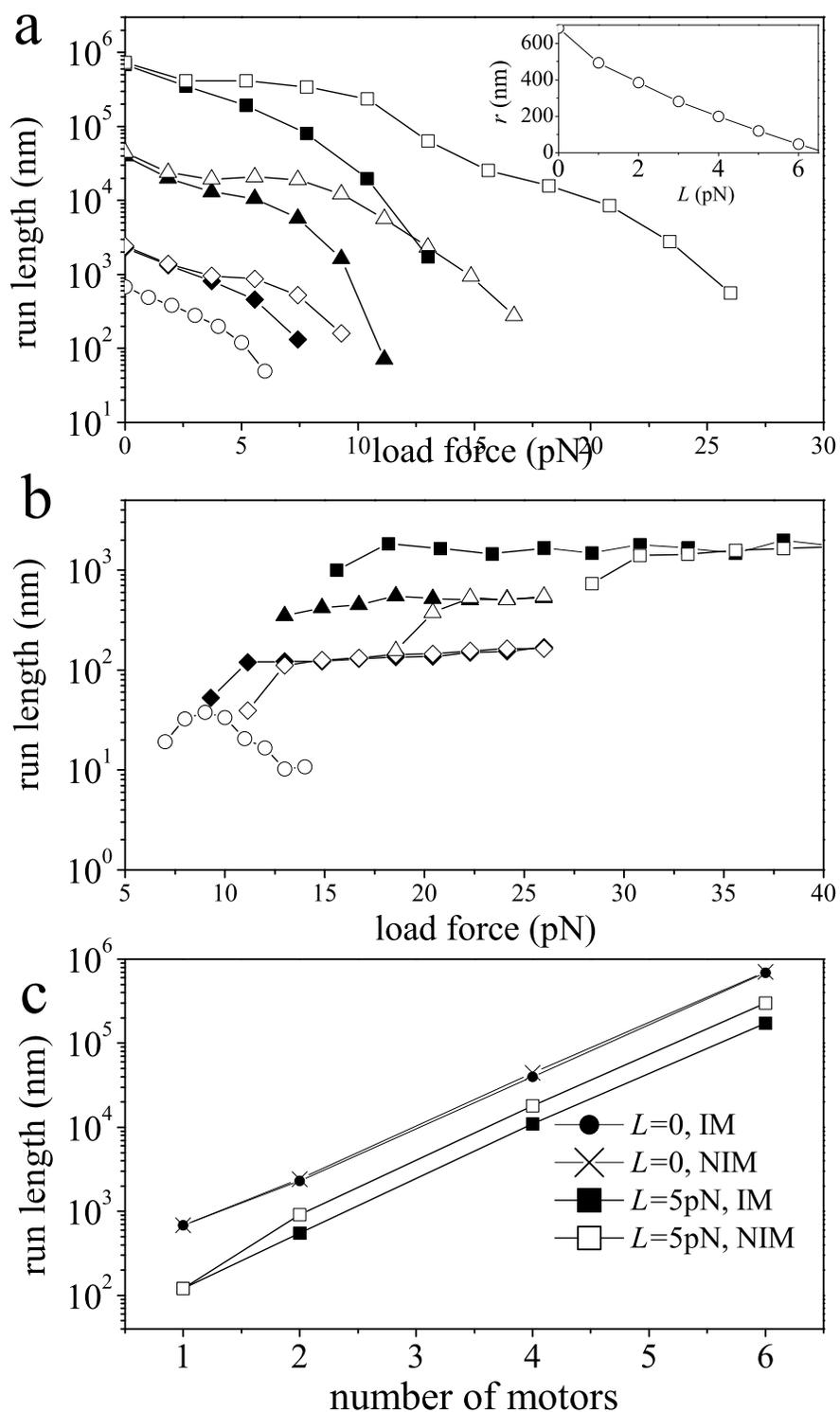}} \caption{\label{f-runl} a) Run lengths as function
of $L$ for $N=1$ (circles), $N=2$ (rhombus), $N=4$ (triangles) and $N=6$ (squares) considering IM (solid symbols) and NIM (open symbols). Results for forward motion of cargo (i.e. relatively small $L$). The inset shows the $N=1$ curve in linear scale
for sort of comparison with experiments in \cite{block2000}. b) Ibid (a) for backward motion of cargo (i.e. relatively large $L$). 
c) Run length as a function of $N$ for two different values of $L$ leading to forward motion of cargo. Results for IM and NIM}
\end{figure}

Figure \ref{f-runl}.a shows $r(L)$ for different values of $N$ considering both IM and NIM in the ranges of $L$
for which we observe forward motion of cargo. For $N=1$ we find results which are compatible with those from experiments
reported in \cite{block2000}. The separation between the curves found for different values of $N$ suggest
an exponential dependence on $N$, in agreement with previous studies \cite{lipowsky09} and with strong dependence
observed in experiments \cite{grossPNAS2007}. The exponential law is confirmed by the results in figure \ref{f-runl}.c. 
We have also found that the mean run times (i.e. the typical time elapsed before all motors become detached) are 
essentially equal for IM and NIM, and show the same exponential dependence on $N$ as $r$ 
(results not shown). Thus, our calculations indicate
that we can approximate $r\simeq v(L) T_0(L) exp( \mu N)$, where $v(L)$ is the mean velocity (which depend on whether 
we consider IM or NIM) and $T_0(L) exp(\mu N)$ is the mean run time, which is essentially the same for IM and NIM. 
(Note that the results in figure \ref{f-runl}.c suggest that $\mu$ is almost independent of the model and of $L$.)
Hence, the fact that, at fixed $N$ and $L$, we observe larger forward processivity for NIM than for IM 
is mainly a consequence of the larger velocities found for NIM. 

In figure \ref{f-runl}.b we show results for $r(L)$ in the ranges of $L$ where we observe backward
motion of the cargo. We first see that, although the values of run length obtained for back motion of cargo 
are much smaller than those for forward motion, they are not fully negligible. Thus, we can clearly speak of 
backward processivity. Again, we find that the mean run times for IM and NIM are indistinguishable (results 
not shown). In this case, however, IM produces larger (backward) processivity than NIM because of the larger negative 
velocities observed for IM. For very large $L$, backward velocities for both models coincide and, consequently, 
the run lengths do so.

\section{Interacting motors on multiple tracks}

Real microtubules provide molecular motors with more than a single track for their advance \cite{hyeon2007,campas2008}.
Motors may surpass one another with relatively small or null interaction if they occupy different tracks.
Microtubules are in fact 13 tubulin protein in perimeter, and, although steric restrictions in the cargo-motor interaction would limit the accessible number of tracks, we still could think on 3 or 4 independent tracks in the filament.
Even more, the possibility of motor advance on parallel independent microtubules could be considered.
Thus, the model with excluded-volume interaction in a single track presented in the previous sections constitutes an extreme simplification that could lead to an overestimation of the effects of interactions. Therefore, in this section we analyze a multi track model in which only the motors that are on the same track undergo excluded-volume interaction,
while those on different tracks do non interact. We consider $N$ motors on $N_T$ tracks. We label each track with an integer number $j_T=1,...N_T$. Motor attachment of detached motors occurs at randomly selected tracks. Once a motor is attached, it never changes its track until it detaches and attaches again. We continue with a one dimensional modeling concerning distances and forces between motors and cargo. No extra assumption is made on the spatial distribution of the different tracks and the distances between them. Actually, the track number $j_T$ of a motor acts effectively as an additional integer coordinate indicating which motors interact with it (those with the same value of $j_T$). Note that for $N_T\to \infty$, we recover the NIM model, since two motors interact with vanishing probability. For finite $N_T$, in contrast, a given pair of motors has always a non null probability of being on the same track.

\begin{figure}[h!]
\centering
\resizebox{\columnwidth}{!}{\includegraphics{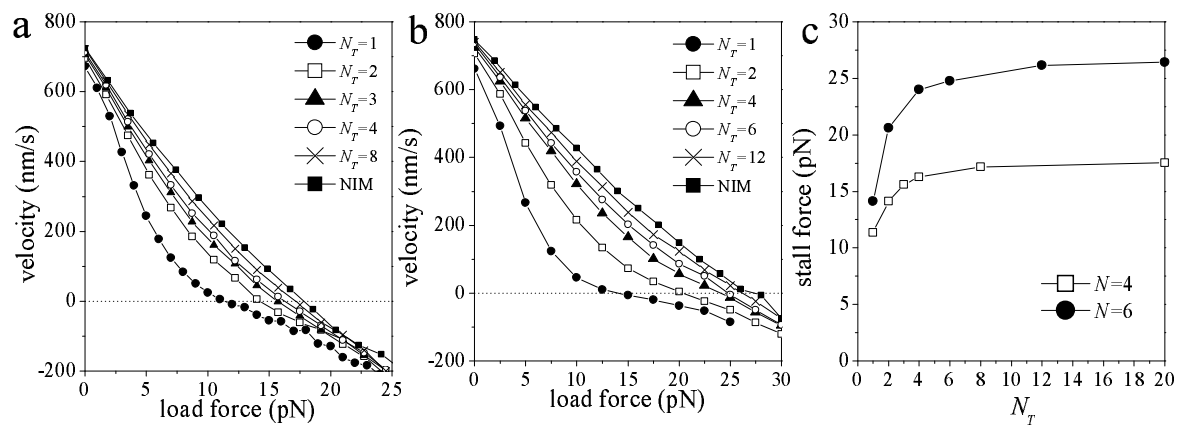}} \caption{\label{f-multitracks} Models for IM
with multiple tracks. a) Mean cargo velocity as function of the load force for $N=4$ considering different
values of $N_T$. b) Ibid (a) for $N=6$. c) Stall force as function of $N_T$ for $N=4$ and $N=6$.}
\end{figure}

In figures \ref{f-multitracks}.a and \ref{f-multitracks}.b we show the force-velocity curves 
for varying $N_T$, for $N=4$ and $N=6$ respectively.
Although the differences between the curves for IM on multiple tracks and the results for NIM decrease with
the number of tracks, they are far from being negligible for reasonable values of $N$ and $N_T$. For instance, for $N=4$
and $L=8.5 pN$ (approximately half of the stall force for NIM), we have that the cargo velocity is of $330 nm/s$ for NIM while it is just $233 nm/s$ for IM with $N_T=3$. Something similar occurs with the results for the stall force shown in \ref{f-multitracks}.c. For instance, for $N=4$, we have that the stall force for $N_T=2$ is $20\%$ smaller than the stall force for NIM, and, for $N_T=3$, we still have a difference of $13\%$ with the value for NIM. Note that the convergence to the NIM model observed for increasing $N_T$ is relatively slow. The force velocity curves for IM with $N_T=2 N$ for systems with $N=4$ and $N=6$ shown in figures \ref{f-multitracks}.a and \ref{f-multitracks}.b are still clearly distinguishable from the curves for NIM.

Similarly to what occurs with the results for the cargo velocity, we have found that the run length for varying $N_T$ presents a smoothly increasing behavior between the limiting values obtained for NIM and IM on a single 
track (results not shown, see figure \ref{f-runl} for limiting values).

In the next section we study the influence of the assumptions for the detachment rate and of the allowance of 
motor back steps. In addition to the consideration of direct interaction between
motors, such modeling details are some of the main differences between our model and other models in the
literature.

\section{Influence of modeling details}

\subsection{Two models for detachment}

Here we analyze the influence of the dependence considered for the motor detachment rate as a function of the load force
$F$ on the dynamics of cargo. Results in previous sections were obtained considering our proposal
$P_{det}(F)$, which is mainly inspired on the more detailed formulation in \cite{grossCB2008}. Here we will compare
them with those coming from the consideration of the pure exponential form $P_{det}^{exp}(F)$ usually found in
the literature \cite{mogilner,lipowsky09,tugofwar}. We consider  IM on a single track and we use
model H for the advance dynamics of single motors.

\begin{figure}[h!]
\centering \resizebox{\columnwidth}{!}{\includegraphics{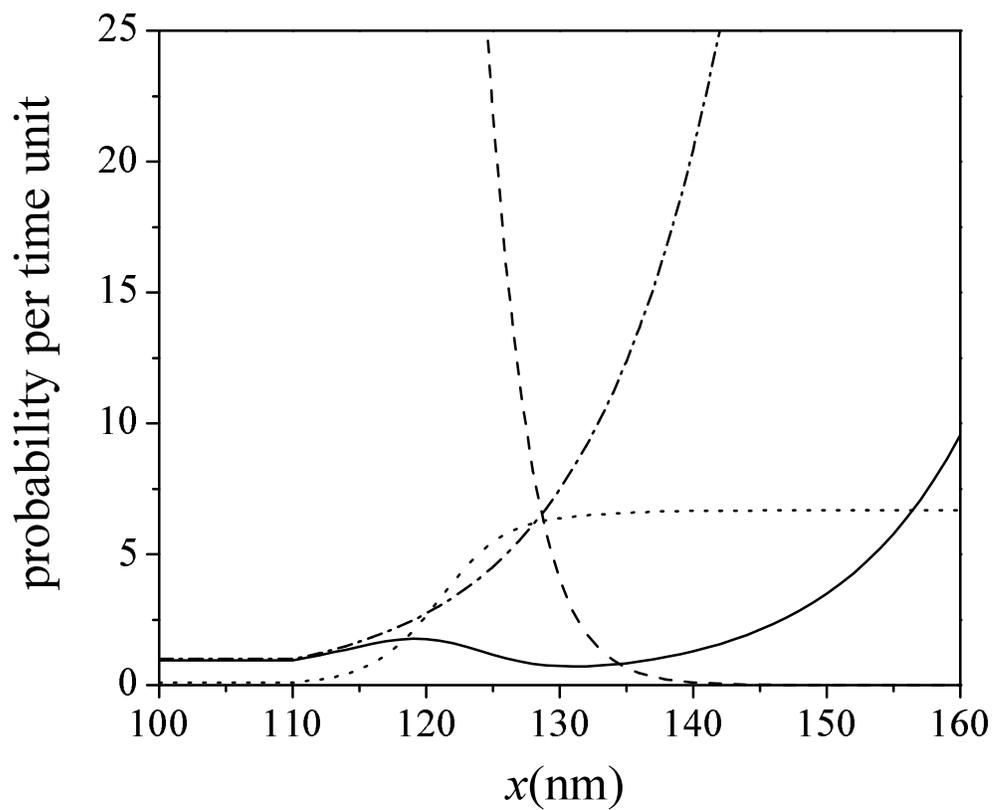}} \caption{\label{fdetachdex} Detaching probability $P_{det}$ (solid line), step forward probability (dashed line) and step back probability (dotted line) as functions of the distance to the cargo for an H-model motor. The dash-dotted line indicates the exponential detaching probability $P_{det}^{exp}$ analyzed in section 5.}
\end{figure}

\begin{figure}[h!]
\centering
\resizebox{\columnwidth}{!}{\includegraphics{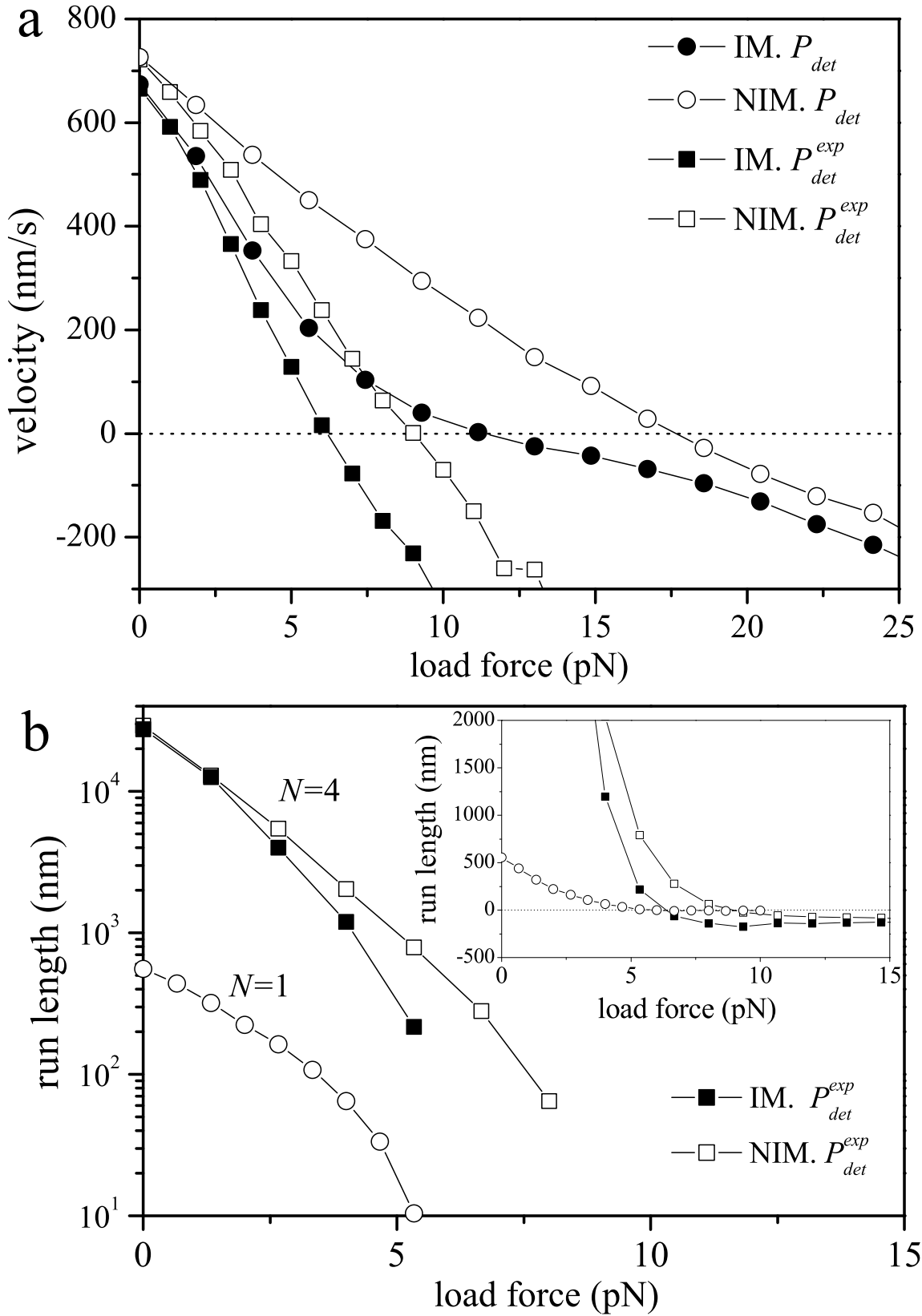}} \caption{\label{f-pdet-exponen} Models with purely
exponential detachment probability $(P_{det}^{exp}(F))$. a) Load-velocity curves for IM and NIM comparing models 
with $P_{det}^{exp}(F)$ and $P_{det}(F)$. Results for $N=4$. b) Run length as function of the load force 
for IM and NIM using $P_{det}^{exp}$. Results for $N=1$ and $N=4$.}
\end{figure}

The difference between the values of $P_{det}$ and $P_{det}^{exp}$ can be appreciated both in figures \ref{fmodels}.d and
\ref{fdetachdex}. While the first compares the explicit formulas as function of $F$,
figure \ref{fdetachdex} shows the dependences of both detachment rates on the distance to the cargo.
The latter figure also enable us to compare the detachment rates with the step forward and step back
probabilities for a motor at a given distance from the cargo. The way in which the values of $P_r$, $P_l$
and the detachment probability compare each other as function of $(x-x_c)$ will clearly influence
the dynamics of motors and, ultimately, the dynamics of cargo. As can be seen, $P_{det}^{exp}$ is larger 
than $P_{det}$ in most of the range of $(x-x_c)$, and it surpasses $P_l$ and $P_r$ at much lower values 
of $F$ than $P_{det}$ does. Hence, in general we could predict that $P_{det}^{exp}$ will lead us to lower 
mean number of attached motors and lower mean number of pulling motors than $P_{det}$,
and thus, to smaller velocities and smaller run lengths. This is actually what we observe
in the results from simulations shown in figure \ref{f-pdet-exponen}. The difference between the results for $P_{det}$ 
and $P_{det}^{exp}$ are smaller at low load forces (typically at $L\lesssim F_d$), since both rates are almost coincident 
in that range. For the case $N=1$, the run lengths found for $P_{det}^{exp}$ are similar to those for $P_{det}$ 
for loads up to the stall force (compare results with those in figure \ref{f-runl}). Thus, taking into account 
the possibility of fine tuning of the parameters, both models for detachment can be considered as equivalent concerning the 
comparison with experimental results in \cite{block2000}. The main differences appear at loads larger than stall, 
for which $P_{det}$ predicts backward processivity while $P_{det}^{exp}$ gives almost vanishing run lengths.
For larger values of $N$ we find that, in the case of forward motion of cargo, the decreasing of the run length with $L$ is much 
faster for $P_{det}^{exp}$ than for $P_{det}$, and that, for the case of back motion of the cargo, $P_{det}^{exp}$ 
lead to very small processivity. Note that in the inset on figure \ref{f-pdet-exponen}.b we have defined $r$ as negative for 
the case of back motion of the cargo, in contrast to what was done in figure \ref{f-runl}, where we considered absolute 
run lengths because of the use of logarithmic scales. In view of our results, the relevance of a relatively small detachment 
rate for enhancing processivity and, in particular, for sustaining back motion of cargo at large loads, is apparent. From a 
more general point of view, our results stress the need for a quite good understanding of the detachment processes and detaching 
rates of single motors in order to obtain exact reliable models on transport by multiple motors. Finally, we want to 
emphasize that, regardless the relatively large differences in the values of velocity and run lengths for both 
models of detachment, the effects of motor interactions are qualitatively the same in both cases. Motor interaction 
always produces a lowering of the velocity, a lowering of the cargo stall force, and a decreasing of the run length. 

\subsection{Models without back steps}

Given a force velocity relation $v(F)$ for individual motors, different Monte Carlo algorithms can be defined providing
compatible stochastic single motor dynamics. In particular, it is possible to consider models without back steps, as was
done in \cite {grossCB2008} and \cite{mogilner}. In such kind of formulations, the motor advances an increment $\Delta x$ with a probability per time unit given by $P_{advance}=v(F)/\Delta x$ for $F<F_0$, while it remain completely stall for $F\ge F_0$.

In this section we analyze the dynamics of a cargo pulled by multiple motors with such kind of single motor modeling
and compare the results with those presented in previous sections obtained with our model that considers both forward and back steps of single motors. The comparison make sense only when the relation
$(P_r(F)-P_l(F))\Delta x=v(F)=P_{advance}(F)\Delta x$ holds. In terms of the forward-backward ratio of jumps $R(F)$ and
the dwell time $\tau_D(F)$ we get
\begin{equation}
P_{advance}=\frac{1}{\tau_D(F)}\frac{R(F)-1}{R(F)+1}.
\end{equation}
Hence, here we compare a model without back steps with such value of $P_{advance}$ for $F<F_0$ (and $P_{advance}=0$
for $F\ge F_0$), to our model with back steps studied in previous sections. The functional forms considered
for $R(F)$ and $\tau_D(F)$ are those indicated in section 2 for model H. The detachment probability
$P_{det}(F)$ is considered in both cases. In the model without back steps, such detachment probability 
has to be handled with care in order to match the average detaching
rates of the model with back steps. The detailed algorithm used is explained in the Appendix.

In figure \ref{f-nostepsback} we show $v$ as function of $L$ for models with and without back steps satisfying the above indicated equivalence relations. We consider both the cases of IM and NIM with $N=4$ and $N=6$. First, it is interesting to observe that back motion of the cargo is quite well described by the model without back steps. This indicates that back motion of the cargo is ruled essentially by the mechanism of detaching of individual motors explained in section 3. The possibility of observing sustained back motion depends on the use of the relatively small probability of detachment given by $P_{det}(F)$ instead of the larger $P_{det}^{exp}(F)$, as was indicated in the previous subsection.

Although the results for models with and without back steps are rather similar, we observe systematic differences both for IM and NIM, and a relevant change in the stall force. It is interesting to note that, while for IM the main differences in the load-velocity curves for models with and without back steps occur for forward motion of cargo, for NIM they arise mostly for backward motion of cargo. This last fact validates the non consideration of motor back steps in the approaches in references \cite{grossCB2008} and \cite{mogilner}, which deal with a team of NIM and
focus their attention on forward motion of cargo.

These effects can be explained as follows. First, in the case of IM for forward motion of cargo we have that
in the model with back steps, actually, back steps are mostly forbidden due to the queue formation.
Thus, the leading motor advances essentially with probability $P_r$ and moves back with very small probability. In contrast, in the model without back steps, the leading motor advances with $P_{advance}=P_r-P_l<P_r$. Hence, the model
with back steps is effectively faster for IM. In contrast, for IM and back motion of the cargo we have that neither back steps nor forward steps occur (the first because they are forbidden by queue effect, the latter because they are rather improbable), and the motion of cargo is ruled by motor detachment which does not depend on whether the model allows or not for motor back steps.

Secondly, for the case of NIM, the differences appear only in the range of $L$ at which back steps probability is appreciable. Clearly, the allowance of back steps contribute to slightly enhance the back velocity of cargo.

The above results depend on our model assumption that considers that interactions between motors at neighbor sites
forbid the forward or backward steps without changing the probabilities $P_r$ and $P_l$. For the IM model, such probabilities actually represent not jumping probabilities, but probabilities of attempting for jumps. We think that this
is actually the best assumption we could make within the framework of our model.

\section{Summary and conclusions}

We have  analyzed the influence of excluded-volume motor-motor interactions on a model for cargo transport driven 
by multiple motors. The model shares a common theoretical framework with other recently proposed in the literature \cite{grossCB2008,mogilner}. It considers a one dimensional Langevin dynamics for cargo, while individual motors evolve following a Monte Carlo algorithm which rules discrete steps on the microtubule, and detaching and attaching as well. The forces between cargo and motors are considered one by one (i.e. not in a mean field approximation), enabling a force sharing analysis. In addition to the effect of direct interaction between motors we have analyzed the relevance of some modeling details, such as the allowance for motor back steps and different prescriptions for the detachment rates.

Our results show that motor interaction leads to appreciable decreasing of the cargo velocity and the cargo stall force
when comparing to the NIM model. The origin of such differences comes from the traffic like phenomena occurring
in the case of IM, which causes a reduction of the number of motors that effectively contribute to pulling from the cargo. Such phenomena were analyzed in terms of spatial distributions of motors, trajectories and force distributions. Concerning the behavior of the cargo stall force with the number of motors $N$, we find that while NIM
model leads to an essentially linear scaling that results valid up to relatively large $N$, IM model shows a rapid saturation for $N\gtrsim 6$. Motor interaction also produces an appreciable reduction of the cargo run length. Such effect is mainly due to the reduction of cargo velocities, as we have checked that the mean time that cargo remains linked to
the microtubule is quite similar for NIM and IM models. This is because, although the mean number of motors that effectively pull from the cargo is higher for NIM that for NIM, the mean number of attached motors is equal for both models.
This is connected to the fact that in the case of IM, motors distribute at shorter distances form cargo. Finally, we also found that the scaling of the run length with $N$ is essentially exponential both for IM and NIM models in agreement
with previous results from other models \cite{klumpp2005,lipowsky09}.

In section 4 we have studied a model with multiple tracks which stands for the possibility of parallel motion of motors in either the same or in different microtubules. In such a model, two motors undergo excluded-volume interaction only when they
are on the same track. The consideration of an increasing number of tracks decreases the effect of interactions on the global properties of cargo transport. However, we have found that the effects of direct interaction of motors remain relevant when realistic numbers of motors and tracks are considered, both on the cargo velocities and on the cargo stall forces.

In section 5 we analyzed two different aspects of the modeling of cargo transport by interacting molecular motors. Namely, the influence of different prescriptions for the detachment rate, and the role of motor back steps. Concerning the first item, we find that different functional dependences of the detachment rate on the load force may lead to quite relevant variations of the results for the cargo velocity, cargo stall force and run length. Thus, a much deeper knowledge of
the way in which single motor detachment rate depends on the applied force would be desirable in order to get reliable models for coupled motors.

Concerning the second item, we have shown that the consideration of motor back steps in models for coupled molecular motors has limited but not completely ignorable relevance. The main differences in the results for models with and without back steps sharing the same force velocity relations for individual motors occur for forward motion of cargo in the case of IM, while for back motion in the case of NIM. In particular, the two kind of models for motor steps differ in their prediction of the cargo stall force, specially in the case of IM. Interestingly, we have found that motor back steps has very little influence on the determination of the transport properties for back motion of cargo. Actually, we have found that back motion of cargo is mainly ruled by motor detachment and that the possibility of an appreciable processivity in back motion depends crucially on the assumptions for the detachment rate.

Our studies provide not only results on systems of interacting motors, but also additional information concerning the case of non interacting motors that can be considered complementary to that on papers as \cite{grossCB2008,mogilner}. We hope that our work may contribute to the general understanding of the rather complex problem of cargo transport by multiple motors.

\section*{Acknowledgments}
This work was possible thanks to the eight-month stay of S.B. at Dpto de F\'{\i}sica de la Materia
Condensada, Universidad de Zaragoza, Spain, supported by CONICET, Argentina. Financial support
of the Spanish MICINN by project FIS2008-01240, coofinanced by FEDER, is also acknowledged.
\section*{Appendix}

\subsection{Parameters defining $\tau_D(F)$ for models H and L.}
\begin{itemize}
\item Model H: $a_1=0.0098s,a_2=0.07s,a_3=0.06(pN)^{-1},a_4=6pN$.
\item Model L: $a_1=0.012 s,a_2=0.18s,a_3=0.6(pN)^{-1},a_4=4.8pN$.
\end{itemize}

\subsection{Monte Carlo algorithm with forward and backward steps for interacting motors (IM)}
At each time step:

\begin{enumerate}
\item For each attached motor:
\begin{itemize}
\item We compute the force $f_i$ depending on $(x_i-x_c)$.
\item We compute the jump probability $dt/\tau_D(f_i)$ and evaluate the possibility of jumping.
\item If the jump is not approved nothing happens and algorithm goes to the next motor.
\item If the jump is approved then
\begin{enumerate}
\item motor $i$ detaches with probability $P_{det}(f_i)$.
\item if motor does not detach, then
\begin{enumerate}
\item it is assigned to perform a forward (backward) jump with probability $P_r$ ($P_l=1-P_r)$.
\item the forward (backward) jump is performed only in case the right (left) site is empty. The new position will
be considered for computing $f_i$  in the next time step. If the right (left) site is not empty nothing happens and
we go to the next motor.
\end{enumerate}
\end{enumerate}
\end{itemize}
\item The cargo advances accordingly to the Langevin dynamics for a time interval $dt$ with $f=\sum_i f_i$. (A standard
stochastic Euler algorithm is considered for the integration.)
\item With probability $\Pi \,dt$, each detached motor attaches to a random empty site located in the interval $|x-x_c|<x_0$.
\end{enumerate}

\subsection{Monte Carlo algorithm with forward and back steps for non interacting motors (NIM)}
Step (i)-(b)-2 in algorithm for IM is suppressed. Jumps assigned in (i)-(b)-1 are always performed.

\subsection{Algorithm without back steps}
In Step (i)-(b)-1 the forward jump is assigned with probability $P_{advance}$, while the backward option is eliminated
both from (i)-(b)-1 and ($i$)-(b)-2.

\subsection{Algorithm using $P_{det}^{exp}(F)$.}
For each motor, the possibility of detachment with probability $P_{det}^{exp}(f_i)$ is evaluated just after
computing $f_i$. If detachment occurs, no evaluation is made of $\tau_D$ and the algorithm goes to the next motor.
Step (i)-(a) is eliminated from the algorithm.

\begin{figure}[h!]
\centering
\resizebox{\columnwidth}{!}{\includegraphics{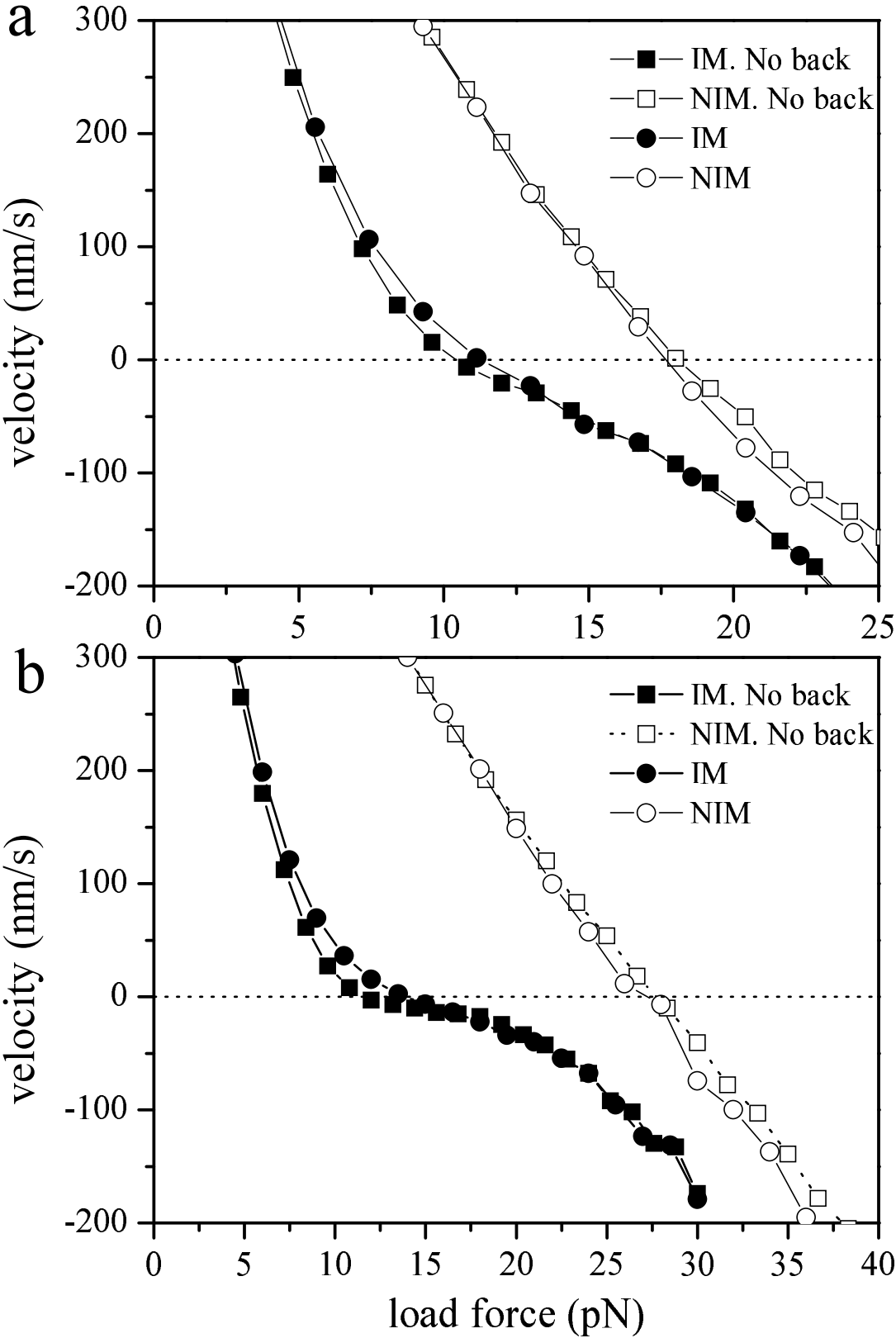}} \caption{\label{f-nostepsback} Comparison of
the models with and without back steps. Mean cargo velocity vs. load force for $N=4$ (a) and
$N=6$ (b).}
\end{figure}

\end{document}